\documentclass[aps,epsf,10pt,showpacs,floatfix,showkeys]{revtex4}
\usepackage{graphicx}
\usepackage{mathrsfs}
\usepackage{amsmath}
\usepackage{amsfonts}
\usepackage{amssymb}
\usepackage{epsfig}
\usepackage{mathrsfs}                                                                                
\usepackage{theorem}
\theorembodyfont{\upshape}

\newcommand{\be}{\begin{equation}}
\newcommand{\ee}{\end{equation}}
\newcommand{\bea}{\begin{eqnarray}}
\newcommand{\eea}{\end{eqnarray}}
\newcommand{\nn}{\nonumber \\}

\input epsf.sty
\usepackage{epsf}
\newlength{\diaght}
\begin{document}
\author{I. Lyberg}
\email{ivar.lyberg@uclouvain.be}
\affiliation{Unit\'e de physique th\'eorique et math\'ematique, Universit\'e catholique de Louvain, 1348 Louvain-La-Neuve, Belgium}
\date{\today}
\title{The free energy of the non-isotropic Ising lattice with Brascamp-Kunz boundary conditions}

\begin{abstract}The free energy of the finite and non-isotropic Ising lattice with Brascamp-Kunz boundary conditions is calculated exactly as a series in the absence of an external magnetic field.
\end{abstract}

\maketitle

\section{Introduction}
\label{int}

The free energy of conformally invariant two dimensional systems have been studied for some time \cite{bcn} \cite{aff} \cite{cardy}. In the case where the system considered is an infinite strip, it is possible to find an exact expression for the free enery per unit length. This has been done in the isotropic case, with equal horizontal and vertical coupling constants \cite{ioh}. This was done by considering a cylindrical Ising lattice of height $\mathcal{M}$ and circumference $2\mathcal{N}$ with so called {\it Brascamp-Kunz} boundary conditions \cite{BK}. In this paper, the corresponding calculation in the case of different vertical and horizontal coupling constants will be done. 

Bl\"ote, Cardy and Nightingale \cite{bcn} wrote the limit of the free energy at criticality per unit length $\lim_{\mathcal{N}\rightarrow \infty}F/2\mathcal{N}$ as
\be \lim_{\mathcal{N}\rightarrow \infty}\frac{F}{2\mathcal{N}}=f\mathcal{M}+f^{\times}+\frac{\Delta}{\mathcal{M}}+...
\ee
where $f$ is the bulk free energy per unit area and $f^{\times}/2$ is the surface free energy. In the isotropic case $\Delta$ is given by
\be \Delta=-\frac{\pi}{24}(c-24h)
\label{ed}
\ee
where $c-24h$ is called the {\it effective} central charge. The effective central charge has been discussed elsewhere, for instance by Izmailian, Priezzhev, Ruelle and Hu \cite{iprh}. For the Ising lattice, the central charge is $c=1/2$, and the allowed values of the conformal weight $h$ are $0$, $1/2$ and $1/16$. It can be seen from equation (17) of ref. \cite{ioh} that the partition function is a multiple of the Virasoro character $\chi_{1/16}=\sqrt{\vartheta_2/2\eta}$ (where the suppressed argument is $i\mathcal{N}/(\mathcal{M}+1)$). Thus $h$ must be 1/16. It will be shown later in this paper that in the isotropic case 
\be \Delta=\frac{\pi}{24}.
\label{delta}
\ee
Since $c=1/2$, (\ref{ed}) and (\ref{delta}) confirm that $h=1/16$.

Equation (\ref{delta}) is only valid if the classical system is rotationally invariant at large distances \cite{bcn} \cite{aff}. In terms of an Ising lattice, this means that the lattice has to be isotropic. If this is not the case, then (\ref{delta}) must be modified by dividing the right hand side by $v$, the ``speed of light''. The speed of light should be obtained from a dispersion relation $\omega\sim vu$, where $\omega$ is the frequency and $u$ is the momentum.

\begin{figure}[bt]
{\epsfig{file=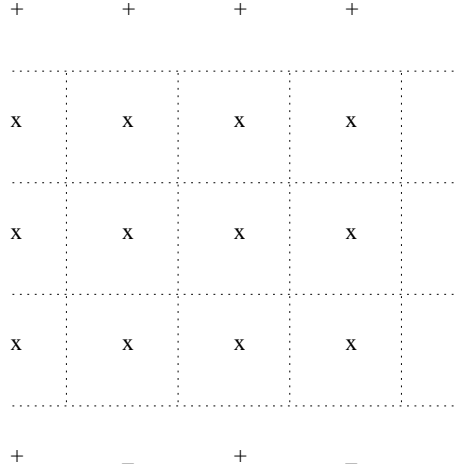, width=6cm}}
\caption{The lattice $\Lambda$ (marked by ``x''; $\mathcal{M}=3$, $\mathcal{N}=2$) and the dual lattice $\Lambda^{*}$ (at intersections of lines).}
\label{fig0}
\end{figure}

The two dimensional cylindrical Ising lattice introduced by Brascamp and Kunz \cite{BK} is the lattice $\Lambda=\{(m,n)~|~1\leq m \leq \mathcal{M},~1\leq n \leq 2\mathcal{N},~(m,2\mathcal{N}+1)=(m,1)\}$ with the following boundary conditions:\newline
{\it (i)} The lattice interacts with a row of fixed, positive spins above it,\newline 
{\it (ii)} The lattice interacts with a row of fixed, alternating spins below it.\newline 
(See fig. \ref{fig0}). The interaction between neighboring spins on $\Lambda$ is $E_1$ in the horizontal direction and $E_2$ in the vertical direction. 
In what follows, the dimensionless parameters
$K_l=\beta E_l~(l=1,~2)$
will be used instead of $E_1$ and $E_2$. Apart from from interactions between nearest neighbors, there is an external magnetic field. 
Thus the Hamiltonian is defined as
\bea \mathcal{E}_{\Lambda}(\sigma,K_1,K_2,H)=-E_1\sum_{j=1}^{\mathcal{M}}\sum_{k=1}^{2\mathcal{N}}\sigma_{j,k}\sigma_{j,k+1}-E_2\sum_{j=0}^{\mathcal{M}}\sum_{k=1}^{2\mathcal{N}}\sigma_{j,k}\sigma_{j+1,k}-\sum_{j=1}^{\mathcal{M}}\sum_{k=1}^{2\mathcal{N}}H(j,k)\sigma_{j,k}
\label{hi}
\eea
where $\sigma_{j,k}=\sigma_{j,k+2\mathcal{N}}=\pm 1$, $\sigma_{0,k}=1$ and $\sigma_{\mathcal{M}+1,k}=(-1)^{k+1}$. The partition function 
\be Z_{\Lambda}(K_1,K_2,H)=\sum_{\sigma \in \{-1,1\}^{\Lambda}}\exp{-\beta\mathcal{E}_{\Lambda}(\sigma,K_1,K_2,H)}
\label{pf}
\ee
has been calculated for the constant external magnetic fields $H\equiv 0$ and $\beta H\equiv i\pi/2$ for which the problem is exactly solvable. Brascamp and Kunz calculated $Z_{\Lambda}(K,K,0)$. $Z_{\Lambda}(K_1,K_2,0)$ and $Z_{\Lambda}(K_1,K_2,i\pi/2)$ were calculated in ref. \cite{L}. $Z_{\Lambda}(K_1,K_2,0)$ is given by
\begin{equation}
Z_{\Lambda}(K_1,K_2,0)=2^{2\mathcal{M}\mathcal{N}}\prod_{j=1}^{\mathcal{N}}\prod_{k=1}^{\mathcal{M}}\{\cosh{2K_1}\cosh{2K_2}-\sinh{2K_1}\cos{\theta_j}-\sinh{2K_2}\cos{\varphi_k}\}\label{p}
\end{equation}
where
\bea &&\theta_j=(2j-1)\pi/2\mathcal{N},~\varphi_k=k\pi/(\mathcal{M}+1).
\label{def1}
\eea

\section{The free energy}
\label{bk}

The partition function (\ref{p}) can be written as
\be Z_{\Lambda}=2^{\mathcal{M}\mathcal{N}}e^{2\mathcal{M}\mathcal{N}\mu(K_1,K_1)}\prod_{j=1}^{\mathcal{N}}\prod_{k=1}^{\mathcal{M}}F(j,k),
\label{z1}
\ee
where 
\bea F(j,k)&:=&4\left(2\sinh^2{\left(\frac{\sinh{2K_2}}{\sinh{2K_1}}\right)^{1/2}\mu(K_1,K_2)}
+\sin^2{\theta_j/2}+\frac{\sinh{2K_2}}{\sinh{2K_1}}\sin^2{\varphi_k/2}\right)
\label{F}
\eea
and the mass $\mu(K_1,K_2)$ is defined as
\be \sinh^2{\mu(K_1,K_2)}:=\frac{1}{4(\sinh{2K_1}\sinh{2K_2})^{1/2}}(\cosh{2K_1}\cosh{2K_2}-\sinh{2K_1}-\sinh{2K_2});
\label{mu}
\ee
in particular
\be \mu(K,K)=\frac{1}{2}\log{\sinh{2K}}.
\label{mui}
\ee
Define $\omega (K_1,K_2;u)$ by the equation
\be \sinh^2{\omega (K_1,K_2;u)}:=2\sinh^2{\left(\frac{\sinh{2K_2}}{\sinh{2K_1}}\right)^{1/2}\mu(K_1,K_2)}+\frac{\sinh{2K_2}}{\sinh{2K_1}}\sin^2{u}.
\label{omega}
\ee
so that
\be F(j,k)=4\left(\sinh^2{\omega (K_1,K_2;\varphi_k/2)}+\sin^2{\theta_j/2}\right).
\label{F2}
\ee
Then, using the identity \cite{gr}
\be \prod_{j=0}^{2\mathcal{N}-1}4\left(\sinh^2{\omega}+\sin^2{\theta_j/2}\right)=4\cosh^2{2\mathcal{N}\omega}
\label{id0}
\ee
one obtains
\be \prod_{j=0}^{2\mathcal{N}-1}F(j+1,0)F(j+1,\mathcal{M}+1)=[4\cosh{2\mathcal{N}\omega (K_1,K_2;0)\cosh{2\mathcal{N}\omega (K_1,K_2;\pi/2)}}]^2.
\label{prod}
\ee
Therefore, using the same argument as in ref. \cite{ioh}, one finds
\be Z_{\Lambda}^2=\frac{2^{2\mathcal{M}\mathcal{N}}e^{4\mathcal{M}\mathcal{N}\mu(K_1,K_1)}}{4\cosh{2\mathcal{N}\omega (K_1,K_2;0)}\cosh{2\mathcal{N}\omega (K_1,K_2;\pi/2)}}\tilde{Z}_{\tilde{\Lambda}}(1/2,0,\mu(K_1,K_2))
\label{z2}
\ee
where $\tilde{\Lambda}$ is a lattice of size $(2(\mathcal{M}+1),2\mathcal{N})$ and the partition function of $\tilde{\Lambda}$ with twisted boundary conditions is given by the equation
\bea (\tilde{Z}_{\tilde{\Lambda}}(\alpha,\beta,\mu))^2&=&\prod_{j=0}^{2\mathcal{N}-1}\prod_{k=0}^{2\mathcal{M}+1}4\Big(2\sinh^2{\mu}+\sin^2{\frac{\pi(j+\alpha)}{2\mathcal{N}}}+\frac{\sinh{2K_2}}{\sinh{2K_1}}\sin^2{\frac{\pi(k+\beta)}{2(\mathcal{M}+1)}}\Big).
\label{tz}
\eea
$(\tilde{Z}_{\tilde{\Lambda}}(1/2,0,\mu(K_1,K_2)))^2$ is given by
\bea (\tilde{Z}_{\tilde{\Lambda}}(1/2,0,\mu(K_1,K_2)))^2&=&\prod_{j=0}^{2\mathcal{N}-1}\prod_{k=0}^{2\mathcal{M}+1}4\Big(
2\sinh^2{\mu(K_1,K_2)}+\sin^2{\theta_{j+1}/2}+\frac{\sinh{2K_2}}{\sinh{2K_1}}\sin^2{\varphi_k/2}\Big)\nn
&=&\prod_{j=0}^{2\mathcal{N}-1}\prod_{k=0}^{2\mathcal{M}+1}F(j+1,k)\nn
&=&\left(\frac{2}{\sinh{2K_1}}\right)^{4\mathcal{N}(\mathcal{M}+1)}\nn
&&\prod_{j=0}^{2\mathcal{N}-1}\prod_{k=0}^{2\mathcal{M}+1}
\bigg(\left(\frac{\sinh{2K_1}}{\sinh{2K_2}}\right)^{1/2}(\cosh{2K_1}\cosh{2K_2}-\sinh{2K_1}-\sinh{2K_2})\nn
&&+2\sinh{2K_1}\sin^2{\theta_{j+1}/2}+2\sinh{2K_2}\sin^2{\varphi_k/2}\bigg).
\label{tz2}
\eea
Define $\tilde{\omega}(K_1,K_2;u)$ by the lattice dispersion relation
\bea \sinh^2{\tilde{\omega}(K_1,K_2;u)}&=&\frac{\sinh{2K_1}}{\sinh{2K_2}}(2\sinh^2{\mu(K_1,K_2)}
+\sin^2{u})\nn
&=&\frac{(\sinh{2K_1})^{1/2}}{2(\sinh{2K_2})^{3/2}}(\cosh{2K_1}\cosh{2K_2}-\sinh{2K_1}-\sinh{2K_2})+\frac{\sinh{2K_1}}{\sinh{2K_2}}\sin^2{u}.
\label{to}
\eea
At criticality it reads 
\be \tilde{\omega}(K_1,K_2;u)\sim \left(\frac{\sinh{2K_1}}{\sinh{2K_2}}\right)^{1/2}u.
\ee
One would thus expect that 
\be v=\left(\frac{\sinh{2K_1}}{\sinh{2K_2}}\right)^{1/2}.
\label{v}
\ee
It will be shown later that this is in fact the case. It follows from (\ref{tz2}) and (\ref{to}) that
\bea (\tilde{Z}_{\tilde{\Lambda}}(1/2,0,\mu(K_1,K_2)))^2&=&\left(\frac{\sinh{2K_2}}{\sinh{2K_1}}\right)^{4\mathcal{N}(\mathcal{M}+1)}\prod_{j=0}^{2\mathcal{N}-1}\prod_{k=0}^{2\mathcal{M}+1}
4(\sinh^2{\tilde{\omega}(K_1,K_2;\theta_{j+1}/2)}+\sin^2{\varphi_k/2}).
\label{zt}
\eea
Using the identity
\be \prod_{k=0}^{2\mathcal{M}+1}4\left(\sinh^2{\omega}+\sin^2{\varphi_k/2}\right)=4\sinh^2{2(\mathcal{M}+1)\omega}
\label{id}
\ee
one obtains from (\ref{zt})
\be \tilde{Z}_{\tilde{\Lambda}}(1/2,0;\mu(K_1,K_2))=\left(\frac{\sinh{2K_2}}{\sinh{2K_1}}\right)^{2\mathcal{N}(\mathcal{M}+1)}\prod_{j=0}^{2\mathcal{N}-1}2\sinh{2(\mathcal{M}+1)\tilde{\omega}\left(K_1,K_2;\theta_j/2\right)}
\label{tz3}
\ee

\section{Asymptotic expansion of the free energy}
Let $(K_1,K_2)=(K^*,K)$ be the curve on which $\mu(K_1,K_2)=0$, or equivalently
\be \sinh{2K_1}\sinh{2K_2}=1.
\label{k1s}
\ee
Let the Taylor expansion of $\tilde{\omega}(K^*,K;u)$ be
\be \tilde{\omega}(K^*,K;u)=\sum_{n=0}^{\infty}\frac{\lambda_{2n}}{(2n)!}u^{2n+1}.
\label{tay}
\ee
In particular 
\be \lambda_0=\left(\frac{\sinh{2K^*}}{\sinh{2K}}\right)^{1/2}=\sinh{2K^*}.
\label{l0}
\ee
Then
\be \tilde{\omega}(K^*,K;u)=\sinh^{-1}{\lambda_0\sin{u}}=\sinh^{-1}{\sinh{2K^*}\sin{u}}
\label{tos}
\ee
and similarly
\be \omega(K^*,K;u)=\sinh^{-1}{\lambda_0^{-1}\sin{u}}=\sinh^{-1}{\sinh{2K}\sin{u}}.
\label{os}
\ee
Thus, on the critical curve, the free energy as obtained from (\ref{z2}) is
\bea F&=&-\log{Z_{\Lambda}}=-\mathcal{M}\mathcal{N}\log{2}-2\mathcal{M}\mathcal{N}\mu (K^*,K^*)+\log{2}\nn
&+&\frac{1}{2}\log{\cosh{2\mathcal{N}\sinh^{-1}{\left(\frac{\sinh{2K}}{\sinh{2K^*}}\right)^{1/2}}}}
-\frac{1}{2}\log{\tilde{Z}_{\tilde{\Lambda}}(1/2,0,\mu(K^*,K))}\nn
&=&-\mathcal{M}\mathcal{N}\log{2}-2\mathcal{M}\mathcal{N}\mu (K^*,K^*)+\log{2}\nn
&+&\frac{1}{2}\log{\cosh{4\mathcal{N}K}}
-\frac{1}{2}\log{\tilde{Z}_{\tilde{\Lambda}}(1/2,0,\mu(K^*,K))}.
\label{fe}
\eea
Clearly 
\bea \log{\tilde{Z}_{\tilde{\Lambda}}(1/2,0,\mu(K^*,K))}&=&2\mathcal{N}(\mathcal{M}+1)\log{\left(\frac{\sinh{2K}}{\sinh{2K^*}}\right)}+2(\mathcal{M}+1)\sum_{j=0}^{2\mathcal{N}-1}\tilde{\omega}\left(K^*,K,\theta_j/2\right)\nn
&&+\sum_{j=0}^{2\mathcal{N}-1}\log{\left(1-\exp{-4(\mathcal{M}+1)\tilde{\omega}\left(K^*,K,\theta_j/2\right)}\right)}.
\label{log2}
\eea
The two sums in (\ref{log2}) can be calculated exactly up to an exponentially small correction $O(e^{-\mathcal{N}})$.

\subsection{Calculation of (\ref{log2})}
The first sum in (\ref{log2}) can be written as a power series using the Euler-Maclaurin summation formula:
\bea 2(\mathcal{M}+1)\sum_{j=0}^{2\mathcal{N}+1}\tilde{\omega}\left(K^*,K,\theta_j/2\right)=\frac{S}{\pi}\int_{0}^{\pi}\tilde{\omega}\left(K^*,K,u\right)du-2\pi\xi \sum_{n=0}^{\infty}\left(\frac{\pi^2\xi}{S}\right)^n\frac{\lambda_{2n}}{(2n)!}\frac{{\rm B}_{2n+2}(1/2)}{2n+2}
\label{sum1}
\eea
where $S=4\mathcal{N}(\mathcal{M}+1)$, $\xi=(\mathcal{M}+1)/\mathcal{N}$ and ${\rm B}_k(1/2)$ is the Bernoulli function ${\rm B}_k(x)$ evaluated at $x=1/2$. It is defined as
\be {\rm B}_p(x):=-\frac{p!}{(2\pi i)^p}\sum_{k\in {\bf Z}\setminus{\{0\}}}k^{-p}e^{2\pi ikx}.
\label{bern}
\ee
It remains to calculate the second sum of (\ref{log2}). 

The second sum in (\ref{log2}) can be written as
\bea &&\sum_{j=0}^{2\mathcal{N}-1}\log{\left(1-e^{-4(\mathcal{M}+1)\tilde{\omega}\left(K^*,K,\theta_j/2\right)}\right)}
=-2\sum_{m=1}^{\infty}\frac{1}{m}\sum_{j=0}^{\mathcal{N}-1}e^{-2m2(\mathcal{M}+1)\tilde{\omega}\left(K^*,K,\theta_j/2\right)}.
\label{log3}
\eea
Let 
$P(p)=\{\pi=(q_1,...,q_{\nu},r_1,...,r_{\nu})~|~q_j,r_j\in {\bf N},~1\leq \nu \leq p,~q_j\neq q_k~{\rm if}~j\neq k,~\sum_{j=1}^{\nu}q_jr_j=p\}$.
The exponential on the right hand side of (\ref{log3}) can be written as
\bea e^{-2m2(\mathcal{M}+1)\tilde{\omega}\left(K^*,K,\theta_j/2\right)}&=&\exp{\left(-2\pi m\lambda_0 \xi (j+1/2)-2\pi m\xi \sum_{p=1}^{\infty}\frac{\lambda_{2p}}{(2p)!}\left(\frac{\pi^2\xi}{S}\right)^{p}(j+1/2)^{2p+1}\right)}\nn
&=&\left(1-2\pi m \xi\sum_{p=1}^{\infty}\left(\frac{\pi^2\xi}{S}\right)^{p}\frac{(j+1/2)^{2p+1}}{(2p)!}\Lambda_{2p}\right)e^{-2\pi m\lambda_0 \xi (j+1/2)}
\label{exp}
\eea
where

\bea
\Lambda_{2p}=(2p)!\sum_{\pi \in P(p)}\left(\prod_{l=1}^{\nu (\pi)}\frac{1}{r_l!}\left(\frac{\lambda_{2q_l}}{(2q_l)!}\right)^{r_l}\right)
(-2\pi m \xi (j+1/2))^{r_1+...+r_{\nu(\pi)}-1}
\label{lam}
\eea

Together, (\ref{log3}), (\ref{exp}) and (\ref{lam}) imply that 
\bea &&\sum_{j=0}^{2\mathcal{N}-1}\log{\left(1-e^{-4(\mathcal{M}+1)\tilde{\omega}\left(K^*,K,\theta_j/2\right)}\right)}=-2\sum_{m=1}^{\infty}\frac{1}{m}\left(\sum_{j=0}^{\mathcal{N}-1}e^{-2\pi m\lambda_0 \xi (j+1/2)}\right)\nn
&&~~~~~~~~~~~~~~~~~~~~~~~~~~~~~~~~~~~~+4\pi \xi \sum_{p=1}^{\infty}\left(\frac{\pi^2\xi}{S}\right)^{p}\frac{1}{(2p)!}\Lambda_{2p}\left(\sum_{m=1}^{\infty}\sum_{j=0}^{\mathcal{N}-1}(j+1/2)^{2p+1}e^{-2\pi m\lambda_0 \xi (j+1/2)}\right).
\label{log4}
\eea
As in ref. \cite{iih}, if, for large $\mathcal{N}$, the finite sum $\sum_{j=0}^{\mathcal{N}-1}$ is replaced by the infinite sum $\sum_{j=0}^{\infty}$ in (\ref{log4}), then equality still holds up to an exponentially small correction $O(e^{-\mathcal{N}})$. Thus
\bea &&\sum_{j=0}^{2\mathcal{N}-1}\log{\left(1-e^{-4(\mathcal{M}+1)\tilde{\omega}\left(K^*,K,\theta_j/2\right)}\right)}=-2\sum_{m=1}^{\infty}\frac{1}{m}\left(\sum_{j=0}^{\infty}e^{-2\pi m\lambda_0 \xi (j+1/2)}\right)\nn
&&~~~~~~~~~~~~~~~~~~~~~~~~~~~+4\pi \xi \sum_{p=1}^{\infty}\left(\frac{\pi^2\xi}{S}\right)^{p}\frac{1}{(2p)!}\Lambda_{2p}\left(\sum_{m=1}^{\infty}\sum_{j=0}^{\infty}(j+1/2)^{2p+1}e^{-2\pi m\lambda_0 \xi (j+1/2)}\right)+O(e^{-\mathcal{N}}).
\label{sum2}
\eea
Combining (\ref{sum1}), (\ref{sum2}) and (\ref{log2}), one obtains 
\bea \log{\tilde{Z}_{\tilde{\Lambda}}(1/2,0,\mu(K^*,K))}&=&2\mathcal{N}(\mathcal{M}+1)\log{\left(\frac{\sinh{2K}}{\sinh{2K^*}}\right)}+\frac{S}{\pi}\int_{0}^{\pi}\tilde{\omega}\left(K^*,K,u\right)du\nn
&-&2\pi\xi \sum_{n=0}^{\infty}\left(\frac{\pi^2\xi}{S}\right)^n\frac{\lambda_{2n}}{(2n)!}\frac{{\rm B}_{2n+2}(1/2)}{2n+2}-2\sum_{m=1}^{\infty}\frac{1}{m}\sum_{j=0}^{\infty}e^{-2\pi m\lambda_0 \xi (j+1/2)}\nn
&+&4\pi \xi \sum_{p=1}^{\infty}\left(\frac{\pi^2\xi}{S}\right)^{p}\frac{1}{(2p)!}\Lambda_{2p}\sum_{m=1}^{\infty}\sum_{j=0}^{\infty}(j+1/2)^{2p+1}e^{-2\pi m\lambda_0 \xi (j+1/2)}+O(e^{-\mathcal{N}})
\label{fe2}
\eea
This expression may be further simplified in terms of elliptic theta functions. However, it is simplest to consider the limit $\mathcal{N}\rightarrow \infty$.

\subsection{The free energy in the limit $\mathcal{N}\rightarrow \infty$}

In the limit $\mathcal{N}\rightarrow \infty$ while $\mathcal{M}$ is fixed, an exact result can be obtained.

Combining (\ref{fe}) and (\ref{fe2}) one obtains
\bea \lim_{\mathcal{N}\rightarrow \infty}\frac{F}{2\mathcal{N}}&=&
-\frac{1}{2}\mathcal{M}\log{2}-\mathcal{M}\mu (K^*,K^*)+K-
\lim_{\mathcal{N}\rightarrow \infty}\frac{1}{4\mathcal{N}}\log{\tilde{Z}_{\tilde{\Lambda}}(1/2,0,\mu(K^*,K))}\nn
&=&-\frac{1}{2}\mathcal{M}\log{2}-(\mathcal{M}+2)\mu (K,K)+K\nn
&-&\lim_{\mathcal{N}\rightarrow \infty}\frac{\mathcal{M}+1}{\pi}\int_{0}^{\pi}\tilde{\omega}\left(K^*,K,u\right)du+\frac{1}{2}\lim_{\mathcal{N}\rightarrow \infty}\frac{1}{\mathcal{N}}\sum_{m=1}^{\infty}\frac{1}{m}\sum_{j=0}^{\infty}e^{-2\pi m\lambda_0 \xi (j+1/2)}\nn
&-&\lim_{\mathcal{N}\rightarrow \infty}\frac{1}{\mathcal{N}}\pi \xi \sum_{p=1}^{\infty}\left(\frac{\pi^2\xi}{S}\right)^{p}\frac{1}{(2p)!}\Lambda_{2p}\sum_{m=1}^{\infty}\sum_{j=0}^{\infty}(j+1/2)^{2p+1}e^{-2\pi m\lambda_0 \xi (j+1/2)}
.
\label{fi}
\eea
The limit of the double sum can be calculated to be
\be \frac{1}{2}\lim_{\mathcal{N}\rightarrow \infty}\frac{1}{\mathcal{N}}\sum_{m=1}^{\infty}\frac{1}{m}\sum_{j=0}^{\infty}e^{-2\pi m\lambda_0 \xi (j+1/2)}=\frac{\pi}{24(\mathcal{M}+1)\lambda_0}
\label{lds}
\ee
Further, by ref. \cite{iih}
\be
\sum_{m=1}^{\infty}\sum_{j=0}^{\infty}(j+1/2)^{2p+1}e^{-2\pi m\lambda_0 \xi (j+1/2)}=\frac{1}{4(p+1)}({\rm B}_{2p+2}(1/2)-{\rm K}_{2p+2}^{1/2,0}(i\lambda_0\xi))
\label{ds2}
\ee
where ${\rm B}_{p}(x)$ is the $p$th Bernoulli function and 
\be {\rm K}_p^{\alpha,\beta}(\tau):=-\frac{p!}{(-2\pi i)^p}\sum_{\substack{m,n\in {\bf Z}\\ (m,n)\neq (0,0)}}\frac{e^{-2\pi i(n\alpha+m\beta)}}{(n+\tau m)^p}
\label{kron}
\ee
is Kronecker's double series.

It can be shown \cite{ioh} that ${\rm K}_{2p}^{1/2,0}(i\xi)$ can be expressed in terms of the elliptic theta functions $\vartheta_2$, $\vartheta_3$ and $\vartheta_4$. It can therefore be shown \cite{ioh} that for small $\xi$
\be {\rm K}_{2p+2}^{1/2,0}(i\lambda_0\xi)={\rm B}_{2p+2}(\lambda_0\xi)^{-2p-2}+O(\xi^{-2p-1})
\label{k}
\ee
where ${\rm B}_n:={\rm B}_n(1)$ is the $n$th Bernoulli number. Since
\be \Lambda_{2p}=\lambda_{2p}+O(\xi)
\ee
for small $\xi$, it follws that
\bea \lim_{\mathcal{N}\rightarrow \infty}\frac{F}{2\mathcal{N}}&=&
-\frac{1}{2}\mathcal{M}\log{2}-(\mathcal{M}+2)\mu (K,K)-\frac{\mathcal{M}+1}{4\pi}\int_{0}^{\pi}\tilde{\omega}\left(K^*,K,u\right)du\nn
&+&K+\frac{\pi}{24(\mathcal{M}+1)\lambda_0}\nn
&+&\sum_{p=1}^{\infty}\left(\frac{\pi}{2(\mathcal{M}+1)}\right)^{2p+1}\frac{{\rm B}_{2p+2}}{(2p)!(2p+2)}
\frac{\lambda_{2p}}{\lambda_0^{2p+2}}.
\label{fi3}
\eea
In particular, one sees that the expected value of $v$ given in (\ref{v}) is correct.

\subsection{The free energy for large $\mathcal{N}$}
We now consider the case of large $\mathcal{N}$, where the partition function is given by (\ref{fe2}). According to ref. \cite{iih},
\be -2\sum_{m=1}^{\infty}\frac{1}{m}\sum_{j=0}^{\infty}e^{-2\pi m\lambda_0 \xi (j+1/2)}=2\sum_{j=0}^{\infty}\log{(1-e^{-2\pi\lambda_0\xi(j+1/2)})}=\log{\frac{\vartheta_{4}(i\lambda_0\xi)}{\eta(i\lambda_0\xi)}}+\pi\lambda_0\xi {\rm B}_2(1/2)
\label{thetaf}
\ee
where $\eta(\tau):=(\vartheta_2(\tau)\vartheta_3(\tau)\vartheta_4(\tau)/2)^{1/3}=e^{i\pi\tau/12}\prod_{n=1}^{\infty}(1-e^{i2\pi\tau n})$. From (\ref{fe2}), (\ref{ds2}) and (\ref{thetaf}) one thus obtains
\bea \log{\tilde{Z}_{\tilde{\Lambda}}(1/2,0,\mu(K^*,K))}&=&2\mathcal{N}(\mathcal{M}+1)\log{\left(\frac{\sinh{2K}}{\sinh{2K^*}}\right)}+\frac{S}{\pi}\int_{0}^{\pi}\tilde{\omega}\left(K^*,K,u\right)du\nn
&-&2\pi\xi \sum_{n=0}^{\infty}\left(\frac{\pi^2\xi}{S}\right)^n\frac{\lambda_{2n}}{(2n)!}\frac{{\rm B}_{2n+2}(1/2)}{2n+2}+\log{\frac{\vartheta_{4}(i\lambda_0\xi)}{\eta(i\lambda_0\xi)}}+\pi\lambda_0\xi {\rm B}_2(1/2)\nn
&+&\pi \xi \sum_{p=1}^{\infty}\left(\frac{\pi^2\xi}{S}\right)^{p}\frac{1}{(2p)!}\Lambda_{2p}\frac{1}{p+1}({\rm B}_{2p+2}(1/2)-{\rm K}_{2p+2}^{1/2,0}(i\lambda_0\xi))+O(e^{-\mathcal{N}})
\label{fe3n}
\eea
It now follows from (\ref{fe}) and (\ref{fe3n}) that 
\bea F&=&-\mathcal{M}\mathcal{N}\log{2}-2\mathcal{M}\mathcal{N}\mu (K^*,K^*)+\frac{1}{2}\log{2}\nn
&+&2\mathcal{N}K
-\mathcal{N}(\mathcal{M}+1)\log{\left(\frac{\sinh{2K}}{\sinh{2K^*}}\right)}-\frac{S}{2\pi}\int_{0}^{\pi}\tilde{\omega}\left(K^*,K,u\right)du\nn
&+&\pi\xi \sum_{n=0}^{\infty}\left(\frac{\pi^2\xi}{S}\right)^n\frac{\lambda_{2n}}{(2n)!}\frac{{\rm B}_{2n+2}(1/2)}{2n+2}
-\frac{1}{2}\log{\frac{\vartheta_{4}(i\lambda_0\xi)}{\eta(i\lambda_0\xi)}}-\frac{1}{2}\pi\lambda_0\xi {\rm B}_2(1/2)\nn
&-&\frac{1}{2}\pi \xi \sum_{p=1}^{\infty}\left(\frac{\pi^2\xi}{S}\right)^{p}\frac{1}{(2p)!}\Lambda_{2p}\frac{1}{p+1}({\rm B}_{2p+2}(1/2)-{\rm K}_{2p+2}^{1/2,0}(i\lambda_0\xi))+O(e^{-\mathcal{N}})
\label{fe4n}
\eea

\acknowledgments

The author thanks Philippe Ruelle and N. Sh. Izmailian for useful discussions. This work was supported by the Belgian Interuniversity Attraction Poles Program P6/02.

\end{document}